\documentclass[acmsmall]{acmart}

 \usepackage{color, colortbl}
 \usepackage{arydshln}
\AtBeginDocument{%
  \providecommand\BibTeX{{%
    \normalfont B\kern-0.5em{\scshape i\kern-0.25em b}\kern-0.8em\TeX}}}
\definecolor{Gray}{gray}{0.9}
\usepackage{algorithmicx}
\usepackage{algpseudocode}
\usepackage{fontawesome}
 \usepackage[linesnumbered,ruled]{algorithm2e}
 \usepackage{pdfpages}
\usepackage[all]{nowidow}
 
\usepackage{subcaption}
 \usepackage{etoolbox}
  \usepackage{booktabs} % For formal tables
\usepackage{float}
\usepackage{amsmath}
\usepackage{diagbox}
\usepackage{color}
\usepackage[utf8x]{inputenc}
 \usepackage{wrapfig}
\usepackage{hyperref}
 
 \usepackage{graphicx}
 \usepackage{graphicx}
\usepackage{color}
\usepackage{xcolor}
\usepackage{url}

\usepackage{graphics}
\usepackage{framed} 
\usepackage{listings}
\usepackage{pifont}
\definecolor{codegreen}{rgb}{0,0.6,0}
\definecolor{codegray}{rgb}{0.5,0.5,0.5}
\definecolor{codepurple}{rgb}{0.58,0,0.82}
\definecolor{backcolour}{rgb}{0.95,0.95,0.92}
\renewcommand{\paragraph}[1]{\vspace{0.05in}\noindent{\bf{#1}.}}

\usepackage{multirow}
\usepackage{booktabs}

\newcolumntype{P}[1]{>{\centering\arraybackslash}p{#1}}
\newcommand{\ignore}[1]{}
\definecolor{tablegray}{gray}{.95}
 
\usepackage{xspace}

\usepackage[T1]{fontenc}

\usepackage{enumerate}

\usepackage[most]{tcolorbox}
\newtcolorbox[%
auto counter]{mybox}[2][]{%
	enhanced jigsaw,
	breakable,
	#1}

\usepackage[normalem]{ulem}

\usepackage{hyperref}
\hypersetup{
    colorlinks = true,
    linkcolor = blue,
    anchorcolor = blue,
    citecolor = blue,
    filecolor = blue,
    urlcolor = blue
}

\newtcolorbox{bbox}[2][]
{
  colframe = #2!25,
  colback  = #2!10,
  enhanced jigsaw,
	breakable,
  #1,
}

 \usepackage{bbding}

\begin{document}

\title{Detecting Scams Using Large Language Models}
%\title{When Good Becomes Evil:\\  Deanonymizing Bluetooth Low Energy Devices via Filter Accept Listing}

%\title{When Good Becomes Evil: Tracking Bluetooth Low Energy Devices via whitelisting}

% When Good Becomes Evil: Bluetooth Address Tracking (BAT) via whitelisting

\author{Liming Jiang}

\affiliation{%
  \institution{Harbin University of Science and Technology}
  \city{Harbin}
  \state{Heilongjiang }
  \country{China} 
}
 
%\author{Anonymous}

%\keywords{Bluetooth Low Energy, Secure Connection Only Mode, Android Security, Downgrade Attack, Enforce Secure Pairing}

%   \settopmatter{printacmref=false} % Removes citation information below abstract
%  \renewcommand\footnotetextcopyrightpermission[1]{} % removes footnote with conference information in first column

%\fancyhead[L]{\fontsize{10}{12} \selectfont  }

%  \author{Yue Zhang, 
%         Zhiqiang Lin
%         % <-this % stops a space
% \IEEEcompsocitemizethanks{\IEEEcompsocthanksitem
%  Yue Zhang and Zhiqiang Lin are with the Computer Science and Engineering, 
%  The Ohio State University,  Columbus Ohio, 43210
%  }% <-this % stops an unwanted space
 
% \thanks{Manuscript received December, 2022; revised December, 2022.}}
 
 \begin{abstract}
     Large Language Models (LLMs) have gained prominence in various applications, including security. This paper explores the utility of LLMs in scam detection, a critical aspect of cybersecurity. Unlike traditional applications, we propose a novel use case for LLMs to identify scams, such as phishing, advance fee fraud, and romance scams. We present notable security applications of LLMs and discuss the unique challenges posed by scams. Specifically, we outline the key steps involved in building an effective scam detector using LLMs, emphasizing data collection, preprocessing, model selection, training, and integration into target systems. Additionally, we conduct a preliminary evaluation using GPT-3.5 and GPT-4 on a duplicated email, highlighting their proficiency in identifying common signs of phishing or scam emails. The results demonstrate the models' effectiveness in recognizing suspicious elements, but we emphasize the need for a comprehensive assessment across various language tasks. The paper concludes by underlining the importance of ongoing refinement and collaboration with cybersecurity experts to adapt to evolving threats.
 \end{abstract}
 
% \IEEEtitleabstractindextext{
% \input{Sections/sec-abstract.tex}

% \begin{IEEEkeywords}
%  Wireless Security, Bluetooth Privacy, Identity Resolution Key (IRK), Side Channel Attacks, Replay Attacks
% \end{IEEEkeywords}}
 
\maketitle

\section{INTRODUCTION}\label{sec:introduction}

A large language model, often referred to as a "large language model" or "LLM," is a type of artificial intelligence (AI) model that is trained on vast amounts of text data to understand and generate human-like text. These models are designed to process and generate natural language text, making them capable of tasks like text generation, text classification, language translation, question answering, and more. They have become increasingly popular in recent years due to their impressive capabilities.

Key characteristics of large language models include: (1) Large language models are characterized by their immense size, often having hundreds of millions to billions of parameters. These parameters are the model's learned weights, which determine its ability to understand and generate text. (2)  Large language models are versatile and can be adapted to perform various natural language processing tasks with minimal task-specific training data. This adaptability has made them popular in a wide range of applications. (3) These models are known for their ability to generate text that is coherent and human-like, often blurring the line between machine-generated and human-generated content.

Recently, LLMs have found various security applications due to their ability to analyze and generate text, which can be instrumental in identifying and addressing security threats. Some notable security applications of LLMs include:

\begin{itemize}
    \item \textbf{Phishing Detection}: LLMs can assist in the detection of phishing emails and websites by analyzing the content and language used in these fraudulent communications. They can flag suspicious messages or URLs based on known phishing patterns and emerging threats.
    \item 
\textbf{Sentiment Analysis: }LLMs can be employed to monitor social media and online forums for discussions related to security incidents or vulnerabilities. Analyzing sentiment can help identify potential threats or vulnerabilities early, allowing for proactive response.
\item \textbf{Threat Intelligence:} Large language models can process vast amounts of unstructured data to extract information about potential threats and vulnerabilities. They can help security analysts by summarizing security reports, identifying emerging threats, and providing context to aid decision-making.

\item \textbf{Malware Analysis: }LLMs can assist in analyzing malware reports, code samples, and security documentation. They can help security professionals understand the behavior of new malware strains and generate signatures or heuristics for detection. 

\item 
\textbf{Vulnerability Assessment: }LLMs can analyze security vulnerability reports, software documentation, and code repositories to assist in identifying and prioritizing software vulnerabilities, which is crucial for timely patching and risk mitigation.
\end{itemize}

\begin{figure}[htbp]
\centerline{\includegraphics[width=4in]{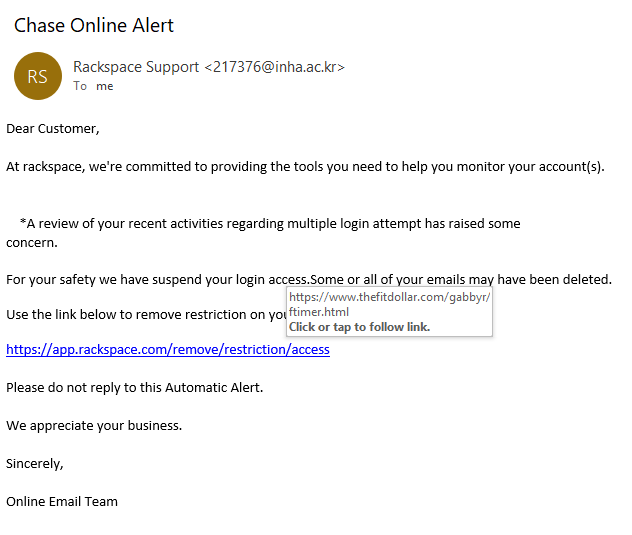}}
\caption{An example of a scam.}
\label{fig1}
\end{figure}

Different from those applications, we propose to use LLM for scam detection. Scams are deceptive practices designed to exploit individuals, organizations, or businesses, often by tricking them into giving away money or personal information. Common scams include phishing, where fraudulent emails and websites impersonate legitimate sources; advance fee fraud, which promises rewards in exchange for upfront payments; romance scams that use fake online dating profiles to solicit money; and investment schemes promising high returns. Tech support scams, online shopping scams, lottery and prize fraud, IRS impersonation, and charity scams are also prevalent. Vigilance and skepticism when encountering unsolicited offers or requests for personal information are crucial to avoid falling victim to these fraudulent schemes. Figure \ref{fig1} shows an example of a scam.

 Building an effective scam detector using LLM involves several key steps. Firstly, it requires comprehensive data collection encompassing various scam types and legitimate content to create a diverse dataset. Data preprocessing is then conducted to clean, standardize, and format the text for training. Labeling is essential, with annotators or crowdsourcing platforms labeling each text as "scam" or "legitimate." The right LLM architecture, like GPT-3 or BERT, is chosen, and fine-tuning on a task-specific dataset is often necessary. Supervised learning is employed for training, followed by rigorous evaluation using metrics like precision, recall, and accuracy. Hyperparameter tuning refines the model, and false positives are analyzed to improve precision. Setting an appropriate confidence threshold for classification is crucial, and finally, integration into the target system ensures real-time scam detection. Collaboration with domain experts and cybersecurity professionals is vital for ongoing refinement and adaptation to emerging threats.

However, the goal of this paper is to introduce and explain a foundational concept while also conducting preliminary evaluations. To achieve this, we duplicated an email and analyzed it using two advanced language models: GPT-3.5 and GPT-4.  The results shown that the email in question is flagged as suspicious due to various red flags, such as an unusual sender address, poor grammar and spelling, suspicious links, an unusual request, lack of personalization, an unusual sender name, and a generic sign-off. 
Two separate analyses, one by GPT-3.5 and another by GPT-4, are presented. Both models identify similar issues in the email, indicating their proficiency in recognizing common signs of phishing or scam emails. However, the paragraph emphasizes that evaluating language models goes beyond a single task, and their effectiveness can vary depending on the complexity of the text.

\begin{figure}[htbp]
\centerline{\includegraphics[width=\textwidth]{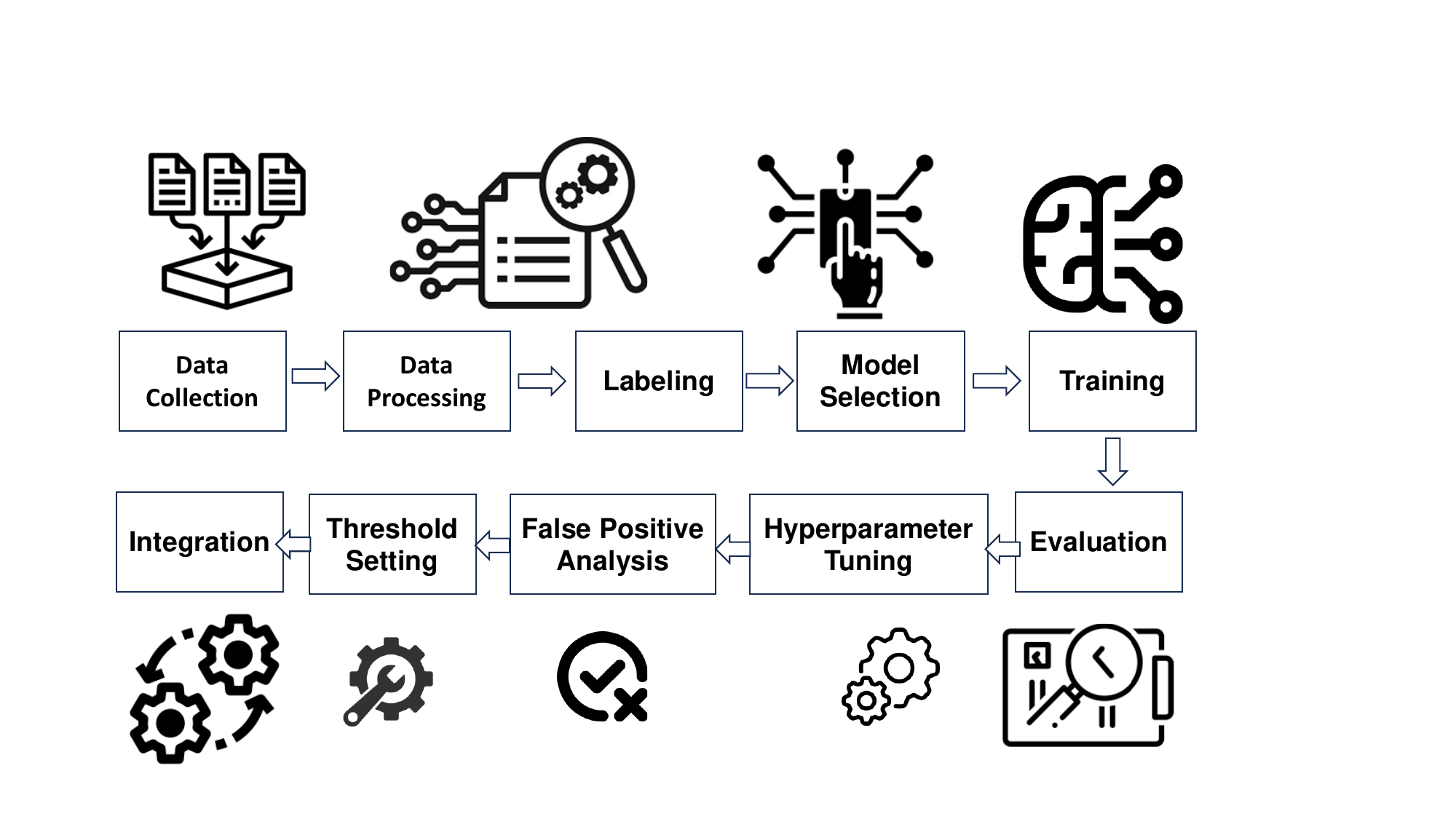}}
\caption{Workflow of the method}
\label{fig2}
\end{figure}

The conclusion is that while both GPT-3.5 and GPT-4 perform well in this specific analysis, a more comprehensive assessment is needed to determine their relative strengths and weaknesses across various natural language understanding and generation tasks.

\section{LITERATURE REVIEW}\label{sec:literature_review}

 Numerous surveys and studies have already been conducted to explore the multifaceted landscape of Large Language Models (LLMs), each with a unique focus and perspective. These comprehensive surveys have touched upon various aspects of LLMs, including their evolution, taxonomy, and applications across different domains such as software engineering and medicine~\cite{chang2023survey,zhao2023survey, wu2023survey,hadi2023survey,wu2023unveiling, bowman2023eight,zhao2023knnicl,fan2023large,hou2023large,thirunavukarasu2023large,clusmann2023future}.

However, this paper takes a distinctive approach by primarily emphasizing the security and privacy facets of LLMs. As we delve into an examination of the existing literature related to this specific focus, we uncover valuable insights into how LLMs, particularly ChatGPT, have the potential to reshape the current cybersecurity landscape. These insights stem from a blend of technical and social considerations, with a particular inclination towards understanding the social dynamics that LLMs introduce in the context of cybersecurity~\cite{caven2023more}.

Moreover, several studies have probed the implications of ChatGPT in cybersecurity, emphasizing its practical applications such as code security and malware detection~\cite{al2023chatgpt,marshall2023effects}. These studies acknowledge the transformative potential of LLMs in enhancing security measures. However, it is worth noting that while they discuss the positive aspects of LLMs, they may not extensively address the potential cybersecurity threats that these models themselves might introduce.
Yifan et al.~\cite{yao2023survey} shed light on how LLMs can both enhance and jeopardize cybersecurity. It underscores the need for continued research in this area to harness the positive aspects of LLMs while mitigating their potential risks and vulnerabilities.

In this landscape, Dhoni et al.~\cite{dhoni2023synergizing} have made notable contributions by showcasing how LLMs can significantly aid security analysts in developing effective solutions against cyber threats. Nevertheless, their work does not delve deeply into the potential vulnerabilities or risks that may arise from the integration of LLMs into cybersecurity practices.
\section{METHODOLOGY}\label{sec:methodology}

Building an effective scam detector using LLMs is an ongoing process that requires constant refinement and adaptation to counter new scamming techniques and emerging threats. Collaboration with domain experts and cybersecurity professionals is essential for a successful scam detection system. As shown in Figure, we list the steps:

\begin{enumerate}
    \item \textbf{Data Collection:}
The first crucial step in creating a scam detector with LLMs is data collection. To build a reliable and robust model, it's essential to gather a diverse and extensive dataset that includes various types of known scams and legitimate content. This dataset should encompass a wide range of scam tactics, languages, and contexts, ensuring that the model can effectively recognize and differentiate between scams and legitimate communications.
\item \textbf{Data Preprocessing:}
Once you've amassed your dataset, the next step is data preprocessing. This involves cleaning and transforming the data to make it suitable for training. Common preprocessing tasks include removing irrelevant information, formatting the text consistently, and addressing issues like spelling errors or inconsistent formatting. Tokenization and stemming may also be applied to standardize the text for analysis.
\item \textbf{Labeling:}
The labeling process is vital for supervised learning. Annotators or crowdsourcing platforms should be employed to label each piece of text in your dataset as either "scam" or "legitimate." It's crucial to ensure that the labeling process maintains a high level of accuracy and consistency. Inter-annotator agreement should be considered to gauge the reliability of the labels.
\item \textbf{Model Selection:}
Choosing the right LLM architecture is a pivotal decision in the development of your scam detection system. There are several LLM options available, including GPT-3, BERT, and custom-built models. The choice depends on your specific requirements and the complexity of the task. Additionally, fine-tuning the selected model on a task-specific dataset is often necessary to optimize its performance for scam detection.
\item \textbf{Training:}
With a labeled dataset and a selected model, it's time to train the LLM for scam detection. Supervised learning techniques are applied, allowing the model to learn from the labeled examples in your dataset. The goal here is for the model to become proficient at classifying text as either a scam or legitimate based on the provided labels.

\item \textbf{Evaluation:}
After training, the model's performance must be rigorously evaluated. Several evaluation metrics, such as precision, recall, F1 score, accuracy, or area under the ROC curve (AUC), can be used to assess its effectiveness. Cross-validation or hold-out validation sets can help ensure that the model performs well across different datasets and scenarios.

\item \textbf{Hyperparameter Tuning:}
To enhance the model's accuracy and robustness, hyperparameter tuning is conducted. Parameters like learning rate, batch size, and model architecture are adjusted and fine-tuned to optimize the model's performance. This iterative process helps in achieving better results.

\item \textbf{False Positive Analysis:}
An essential step in the development of a scam detector is the analysis of false positives generated by the model. By carefully examining and addressing these false alarms, the model can be refined to reduce the number of legitimate communications mistakenly classified as scams. This iterative process is vital for improving the detector's precision.

\item \textbf{Threshold Setting:}
Determining an appropriate confidence threshold for classification is a critical decision. Text with a confidence score above this threshold is classified as a scam, while those below it are considered legitimate. The threshold should strike a balance between minimizing false positives and false negatives, aligning with your desired detection strategy.

\item \textbf{Integration:}
Once the model is trained and fine-tuned, it needs to be integrated into your application, platform, or system. This integration should enable the model to efficiently process incoming text data, make predictions in real-time, and take appropriate actions when scams are detected.

\end{enumerate}

\section{PRELIMINARY ASSESSMENT}

In the context of this paper, our primary objective is to introduce and elucidate a foundational concept while also performing preliminary evaluations. To accomplish this, we have taken the step of duplicating the email mentioned earlier and have subsequently fed this duplicated content into two advanced language models, namely GPT-3.5 and GPT-4. The ensuing feedback received from these models provides valuable insights and assessments, which we shall explore and expound upon in greater detail within the subsequent sections of this paper. The email for test is the email displayed in Figure \ref{fig1}. 

\vspace{2mm}

\noindent \textbf{GPT3.5.} This email is very likely to be a scam. There are several red flags indicating that it's not legitimate:

\begin{itemize}
    \item \textbf{Email Address: } The email is sent from an unusual email address, "217376@inha.ac.kr," which does not appear to be related to Rackspace, a reputable hosting and cloud computing company.

\item \textbf{Poor Grammar and Spelling}: The email contains several grammatical errors and awkward phrasing, such as "we have suspend your login access" and "Some or all of your emails may have been deleted." Legitimate companies typically have professional communication.

\item \textbf{Suspicious Links}: The email contains a link to a website called "wwwthefitdollar.com/gabbyr," which is not related to Rackspace or any legitimate financial institution. Scammers often use deceptive links to phishing websites to steal your personal information.

\item \textbf{Unusual Request}: The email asks you to click on a link to remove restrictions on your account, which is a common tactic used by scammers to trick recipients into providing sensitive information.

\item \textbf{Lack of Personalization}: The email does not address you by name, which is unusual for legitimate businesses that often personalize their communications.

\item \textbf{Unusual Sender Name}: The sender's name appears as "Rackspace Support," but the email address does not match this name, further indicating that it may be a scam.

\item \textbf{Generic Sign-off}: The email closes with a generic sign-off, "Sincerely, Online Email Team," without providing any specific contact information or phone number.
\end{itemize}

In summary, this email displays numerous signs of a phishing attempt or scam. It's crucial not to click on any links or provide any personal information in response to such emails. Instead, if you have concerns about your account, always contact the company directly through their official website or customer support phone number to verify the information.

\noindent \textbf{GPT4.} The email you received contains several signs that suggest it could be a scam:
\begin{itemize}
    \item \textbf{Suspicious Sender Address}: The email is from a sender with an address ending in "@inha.ac.kr", which is not a domain typically associated with Rackspace. Legitimate emails from companies usually come from their own domain (e.g., @rackspace.com).
    \item  \textbf{Unusual Links:} The links provided in the email (like "wwwthefitdollar.com" and a strangely formatted Rackspace link) do not seem to be legitimate Rackspace domains. Scammers often use misleading URLs that resemble legitimate ones to trick recipients into clicking them.
    \item \textbf{Grammar and Spelling Errors:} The email contains several grammatical and spelling errors ("suspend" instead of "suspended", "your login access.Some", etc.), which are not expected in official communication from a reputable company like Rackspace.
   \item \textbf{Sense of Urgency and Fear}: The email creates a sense of urgency by stating that your login access has been suspended and that your emails may have been deleted. Scammers often use these tactics to prompt a quick, emotional response.

\item \textbf{Request Not to Reply:} The instruction not to reply to the email is suspicious. Legitimate companies usually provide a way for you to contact them for verification.

Generic Salutation: The email addresses you as "Dear Customer" instead of using your name. Legitimate emails from your service providers usually address you by name.

\end{itemize}

Considering these factors, it is highly advisable not to click on any links or provide any personal information. You should directly contact Rackspace through their official website or customer support channels to verify the authenticity of the email

\vspace{2mm}

\noindent \textbf{Result Analysis:} While both GPT-3.5 and GPT-4 appear to perform similarly in their assessment of the email, it's important to note that the evaluation of language models goes beyond just a single task or context. The effectiveness of these models can vary depending on the complexity and nuances of the text being analyzed.
In this particular instance, where the email is relatively straightforward and its red flags are apparent, both GPT-3.5 and GPT-4 are able to identify the key issues with the email, such as the suspicious sender address, unusual links, grammar and spelling errors, the sense of urgency, and the generic salutation. Their assessments align closely, demonstrating that both models are proficient in recognizing common signs of phishing or scam emails.
However, when assessing the overall performance of language models like GPT-3.5 and GPT-4, it's essential to consider their capabilities across a wide range of tasks and domains. These models are often evaluated on their ability to understand and generate human-like text, provide contextually relevant responses, and handle more intricate language tasks.
Furthermore, the quality of results can also depend on the training data, fine-tuning methods, and the specific versions of these models. It's advisable to conduct comprehensive evaluations and comparisons on various tasks and datasets to determine their relative strengths and weaknesses accurately.
In conclusion, while both GPT-3.5 and GPT-4 demonstrate comparable proficiency in analyzing this particular email for signs of phishing, a more in-depth and comprehensive assessment would be needed to definitively determine which model excels in a broader spectrum of natural language understanding and generation tasks.

\section*{CONCLUSION}
In conclusion, this paper has explored the potential application of large language models (LLMs) in the critical domain of scam detection within the cybersecurity landscape. LLMs, with their remarkable adaptability and natural language understanding abilities, offer a promising avenue for enhancing the identification and mitigation of fraudulent schemes that pose a significant threat to individuals and organizations alike. Through an analysis of various security applications of LLMs and a detailed description of the steps involved in building an effective scam detector, this paper has laid the groundwork for harnessing the power of LLMs in combating online scams.

Furthermore, the preliminary evaluation conducted using advanced language models, GPT-3.5 and GPT-4, has demonstrated their proficiency in recognizing common red flags associated with scam emails. However, it is crucial to acknowledge that evaluating LLMs extends beyond a single task, and their effectiveness may vary across different linguistic complexities and contexts. 
Moving forward, additional research is necessary to comprehensively assess the strengths and weaknesses of LLMs in various scam detection scenarios. Collaboration with domain experts and continuous adaptation to emerging threats are vital aspects of this endeavor. By further refining and optimizing LLMs for scam detection, we can bolster online security measures, thereby safeguarding individuals and organizations from financial and personal harm caused by deceptive practices. Ultimately, the integration of LLMs into cybersecurity protocols holds the potential to significantly enhance our ability to combat online scams in an ever-evolving digital landscape.

 \bibliographystyle{IEEEtran}
 
% Loading bibliography database
\bibliography{reference}

\end{document}